# Two-dimensional Penta-Pt$_2$N$_4$: an ideal material for nanoelectronics

Zhao Liu[1*], Haidi Wang[1*], Jiuyu Sun[1], Z. F. Wang[1,2] and Jinlong Yang[1,3]

[1]Hefei National Laboratory for Physical Sciences at the Microscale, University of Science and Technology of China, Hefei, Anhui 230026, China

[2]CAS Key Laboratory of Strongly-Coupled Quantum Matter Physics, University of Science and Technology of China, Hefei, Anhui 230026, China

[3]Synergetic Innovation Center of Quantum Information and Quantum Physics, University of Science and Technology of China, Hefei, Anhui 230026, China

[*]These authors contributed equally to this work.

Correspondence and requests for materials should be addressed to Z.F.W. (email: zfwang15@ustc.edu.cn) or J.Y. (email: jlyang@ustc.edu.cn)

**Since the discovery of graphene, two-dimensional (2D) materials have paved a new routine for designing high-performance nanoelectronic devices. To facilitate the device applications, there are three key requirements for a material: sizeable band gap, high carrier mobility and robust environmental stability. However, for the hottest 2D materials studied in recent years, graphene is gapless, transition metal dichalcogenides have low carrier mobility, and black phosphorene is environmentally sensitive. So far, these three characteristics can seldom be satisfied in one single material. Therefore, it becomes a great challenge for finding an ideal 2D material to overcome such a limitation. In this work, we theoretically predict a novel planar 2D material: Penta-Pt$_2$N$_4$, which are designed by the fantastic Cario pentagonal tiling as well as the rare nitrogen double bond. Most significantly, the 2D Penta-Pt$_2$N$_4$ exhibits excellent intrinsic**

**properties, including large direct band gap up to 1.51 eV, high carrier mobility up to $10^5$ cm$^2$·V$^{-1}$ s$^{-1}$, giant Young's module up to 0.70 TPa, and robust dynamic, thermal and chemical stability. Moreover, Penta-Pt$_2$N$_4$ is a global minimal with PtN$_2$ stoichiometry in 2D, so we also propose a CVD/MBE scheme to enable its experimental synthesis. We envision that the 2D Penta-Pt$_2$N$_4$ may find a wide application for future nanoelectronics.**

**Introduction**

The past decade has witnessed a huge progress in the fundamental and application research for the 2D materials[1-3]. As a new material category compared to the 3D materials, the 2D materials have attracted incredible interest from the nanoelectronic field[4], since the atomic-scale thickness can greatly reduce the conventional short-channel effect. The rise of 2D materials began with graphene[5,6], and it's expected to be a perfect material to substitute the silicon in semiconductor industry. However, after the early days of studies, the gapless feature indicates that graphene is less desirable to fulfill this high expectation[7,8]. Hence, the research is expanded into other 2D materials, predominantly transition metal dichalcogenides (TMDs), because of its sizeable bandgap in the range of 1-2 eV. However, compared to graphene, the low carrier mobility in TMDs inevitably limit the device performance speed and power efficiency[9,10]. Later on, the discovery of black phosphorene brings a new ray to the research. It inherits the merits of graphene and TMDs, displaying a sizeable band gap and high carrier mobility. Consequently, the high-performance devices are obtained in a short period of time. Unfortunately, black phosphorene is environmentally sensitive, it can react with water or oxygen, and degrade upon exposure to air as well as light[11-15], which severely reduces its device quality and performance. Therefore, a straightforward question is in front of us, can we find a 2D material with sizeable band gap, high carrier mobility and environmental stability simultaneously?

To answer this question, a feasible way to design such an ideal material is needed. One knows that if the $\pi/\pi^*$ degeneracy is broken with a sizeable band gap, graphene will be the best choice. Following this line, instead of the hexagon in graphene, we turn to the pentagon. Due to the rule of Cario pentagonal tiling, it's possible to construct a 2D lattice full of pentagons, but without $\pi/\pi^*$ degeneracy. Hence, it may be a promising material candidate. However, to realize such a proposal, the following difficulties must be overcome firstly. It's well known that pentagon doesn't prefer to connect with each other in real materials. In fullerene, pentagons and hexagons are distributed alternately, because it will cost more energy to realize pentagon-pentagon than pentagon-hexagon rings[16,17]. This is also true in 2D planar materials[18-20] in which $sp^2$ hybridization is dominant, namely, pentagons prefer to connect with other polygons (e.g. heptagon). On the other hand, all Penta-based materials are quasi-planar[21-27] and no planar structures have been report yet. Physically, the non-planar structure can trace back to the feature of $sp^3$ hybridization which favors buckling to lower its total energy. Therefore, in order to design a Cario pentagonal tiling 2D planar material with novel electronic properties, $sp^2$ and $sp^3$ hybridization should both be avoided. However, this requirement can hardly be satisfied for materials with only main group elements. As an alternative way, we utilize the concept in coordination chemistry, and use transition metal pernitrides[28-42] with nitrogen double bond (N=N) as the basic building block to realize our proposal, considering the fact that *s* and *p* orbital can hardly be hybridized in N=N (see SI for details).

In this work, we report density functional theory (DFT) calculations for the first Cario pentagonal tiling 2D planar material: Penta-$Pt_2N_4$. The interplay between its novel geometric topology and exotic chemical compositions endows excellent material properties for such one-atomic-thick structure, including large direct band gap up to 1.51 eV, high carrier mobility up to $10^5$ $cm^2 \cdot V^{-1} s^{-1}$, giant Young's module up to 0.70 TPa, and robust environmental stability. Furthermore, a possible synthesis method based on the CVD/MEB technique is also proposed to facilitate its experimental realization. Our results not only combine the Cario pentagonal tiling and rare nitrogen double bond in a 2D planar structure, but also demonstrate an ideal 2D material with high comprehensive mechanical and electronic properties for future nanoelectronics.

**Results**

**Structure and stability.** As shown in Fig. 1(a), the Penta-$Pt_2N_4$ has a space group of P4/mbm, and its lattice constant is 4.80 Å. There are four pentagon rings in a unit cell, showing the shape of Cario pentagonal tiling. Each Pt atom is in the square planar crystal field formed by four nitrogen dimers. The Pt-N bond is 2.00 Å, and the N=N bond is 1.26 Å. Before showing the excellent intrinsic properties of Penta-Pt$_2$N$_4$, we try to study its stability firstly.

To confirm the dynamic stability of Penta-Pt$_2$N$_4$, its phonon spectrum is calculated. As shown in Fig. 1(b), the absence of imaginary modes in the entire Brillouin zone (BZ) indicates it to be dynamic stable. Near the Γ point, the in-plane longitudinal and transverse modes have a linear dispersion, while the out-of-plane modes have a quadratic dispersion. All these features are consistent with the other 2D materials. Additionally, the two highest optical modes (~1387cm$^{-1}$), contributed by the N=N, are separated from the other optical modes by a large phonon gap of ~500 cm$^{-1}$. These two modes are Raman-active, representing the characteristic stretching vibration of N=N, and also comparable to the other N=N system[40-42].

To confirm the thermal stability of Penta-Pt$_2$N$_4$, its cohesive energy is calculated as

$$E_{coh} = (2E_{Pt} + 4E_N - E_{Pt_2N_4})/6 \qquad (1)$$

where $E_{Pt}$, $E_N$ and $E_{Pt_2N_4}$ are DFT energy of single Pt atom in the bulk, half energy of N$_2$ and energy of Pt$_2$N$_4$ monolayer, respectively. The calculated cohesive energy of Penta-Pt$_2$N$_4$ is 5.04 eV per atom, which is much higher than that for the theoretically predicted Be$_2$C$_5$[43] (4.58 eV) and experimentally synthesized phosphorene[44,45] (3.61 eV) and silicene[46,47] (3.71 eV). Moreover, AIMD simulation is performed at 1200 K. As shown in Fig. 1(c), the fluctuation of total energy is within a small range of 0.2 eV, illustrating the thermal stability at high temperature. The last snapshot of AIMD simulation is shown in Fig. 1(d), and the average nitrogen bond is ~1.23 Å, which is still within the double bond region. Such a high thermal stability indicates that there will be a large energy barrier between Penta-$Pt_2N_4$ and the other local minimum structures on the potential energy surface of stoichiometric $PtN_2$. To identify this point, we have performed a global search for the lowest energy structure of $[PtN_2]_n$ (n=1, 2, 3, 4) (see Method). As shown in Fig. 1(e), four stable structures of $[PtN_2]_n$ are found. Comparing all these structures, one can see the average energy of Penta-$Pt_2N_4$ is 41, 85 and 88 meV/atom lower than that of β-$Pt_3N_6$, γ-$PtN_2$ and δ-$Pt_4N_8$, respectively. Therefore, Penta-$Pt_2N_4$ is a global minimum structure under this stoichiometric ratio.

To confirm the chemical stability of Penta-$Pt_2N_4$, its chemical adsorption is calculated. Five most common gases, including $CO_2$, $H_2$, $N_2$, $O_2$ and $H_2O$, are used to mimic the air environment. The corresponding stable adsorption configurations (see also SI) and adsorption distances are shown in Fig. 2(a)-(e), respectively. The adsorption energy ($E_a$) is calculated as

$$E_a = E_{Mole-Pt_2N_4} - E_{Mole} - E_{Pt_2N_4} \qquad (2)$$

As shown in Table 1, all adsorption energies are within the range of van der Waals interaction. Here, $O_2$ has a slightly larger $E_a$ than the other gases. This is consistent with its shorter adsorption distance [Fig. 2(d)], which can be explained as the higher chemical activity of $O_2$. Similar tendency has also been observed in graphene[49,50]. Therefore, based on the analysis of adsorption energy and adsorption distance, one notices that all these gases are physically absorbed on Penta-$Pt_2N_4$, demonstrating a highly environmental stability under the air exposure.

**Mechanical properties.** After confirming the dynamic, thermal and chemical stability of Penta-$Pt_2N_4$, its outstanding inherent properties are discussed in the following part. The elastic constants of Penta-$Pt_2N_4$ (see Table S1) satisfy the mechanical stability criteria for tetragonal 2D materials ($C_{11}C_{22}-C_{12}^2>0$, $C_{66}>0$), indicating it to be mechanically stable. Because of the large characteristic vibration frequency of N=N, its Young's modulus is as large as 0.70 TPa (224 N m$^{-1}$) along the *x* direction, as shown in Fig. 2(f). This value is also comparable to that in graphene (see Table 3), but much larger than that in the other 2D transition metal pernitrides with N-N single bond[48]. For the Poisson's ratio of Penta-$Pt_2N_4$, it varies spatially with a maximal value of 0.47 along diagonal direction and a minimal value of 0.18 along *x* direction, as shown in Fig. 2(g). To further explore the ideal tensile strength and critical strain of Penta-$Pt_2N_4$, an in-plane uniaxial tensile strain is applied. As shown in Fig. 2(h), the ideal strength is 36 GPa and 73 GPa, and the critical strain is 12.5 % and 10.4 % along *x* and diagonal direction, respectively.

**Electronic properties.** Besides the mechanical properties, the electronic properties of Penta-$Pt_2N_4$ are also investigated. The PBE band structure of Penta-$Pt_2N_4$ is shown in Fig. 3(a). One can see there is a direct band gap of 0.07 eV at M point. Including the spin-orbit coupling (SOC), this band gap can be further increased to 0.33 eV, as shown in Fig. 3(b). Generally, the PBE will underestimate the band gap, so HSE band structures are also calculated. As shown in Fig. 3(c) and 3(d), HSE increases the direct band gap to 1.10 eV without SOC, and to 1.17 eV with SOC. A detailed comparison between PBE and HSE band structure is shown in the SI. It's clear that HSE has a rigid effect on the band structure, namely, it just corrects the band energy but without changing the dispersion near band edge. Moreover, GW calculation can further increase the direct band gap to ~1.51 eV with SOC (see Methods).

The sizeable band gap indicates Penta-$Pt_2N_4$ to be a good material candidate for nanoelectronics. For electronic applications, another key factor is the carrier mobility. As listed in Table 2, the carrier mobility is calculated on the PBE+SOC level. One can see both electron and hole have a high mobility, which is much large than that in black phosphorene at room temperature. Moreover, the electron mobility along $x$ direction can reach $1.1 \times 10^5$ $cm^2 \cdot V^{-1} s^{-1}$, which is even comparable to the value in graphene, as listed in Table 3. Additionally, strong anisotropy is found for both carries, and the mobility along $x$ direction is one order larger than that along diagonal direction. The high carrier mobility of Penta-$Pt_2N_4$ can be understood from two aspects. First, the large elastic modulus $C_{2D}$, which is originated from the nature of N=N. Second, the small band edge deformation potential, which is originated from the nature of band edge states. As shown in Fig. 3(e)-(h), the band edge partial charge densities have an out-of-plane shape, which will be less sensitive to the in-plane lattice deformation. Therefore, it has a small band edge deformation potential ($E_l$), as listed in Table 2.

Based on the calculated mechanical and electronic properties of Penta-Pt$_2$N$_4$, Table 3 makes a summary and compares them with graphene, MoS$_2$ and black phosphorene. From this direct comparison, the high comprehensive merits of our proposed Penta-Pt$_2$N$_4$ can be seen clearly. Therefore, Penta-Pt$_2$N$_4$ will be an ideal 2D material for nanoelectronics. Since the novel geometric topology and exotic chemical compositions are closely linked to the excellent properties of Penta-Pt$_2$N$_4$, we also try to give a physical understanding about the Cario pentagonal tiling and N=N double bond from the view of metal-ligand coupling.

**Nature of N=N.** To analysis the interaction between Pt and N$_2$, the orbital projected band structures of Penta-Pt$_2$N$_4$ are shown in Fig. 4(a)-(f). Here, the *s* orbital is omitted because it is far away from the Fermi-level (below −10 eV). Due to the symmetry of crystal field, $3\sigma_g$, $3\sigma_u^*$, $1\pi_u$ and $1\pi_g^*$ (formed by $p_x$ and $p_y$ orbitals) of N=N can couple with $d_{x^2-y^2}$, $d_{xy}$ and $d_{z^2}$ orbital of Pt atoms [see state labeled A and B in Fig. S9 (b)], while $1\pi_u$ and $1\pi_g^*$ (formed by $p_z$ orbital) of N=N can couple with $d_{xz}$ and $d_{yz}$ orbital of Pt atoms [see state labeled C and D in Fig. S9 (b)]. Since the coupling strength is proportional to the relative band width of these *d* orbitals, one can see the strongest coupling is between $1\pi_g^*$ of N=N and $d_{x^2-y^2}$ orbital of Pt atoms, which pushes the main component of $d_{x^2-y^2}$ to conduction band and $1\pi_g^*$ to valence band. This is also consistent with the crystal orbital Hamilton population (COHP) analysis for N=N dimer, as shown in Fig. 4(g). The bonding states of $1\pi_u$ and $3\sigma_g$ are far away from the Fermi-level (below −5 eV), while the antibonding states of $1\pi_g^*$ (formed by $p_x$ and $p_y$ orbitals) and $1\pi_g^*$ (formed by $p_z$ orbital) are just below and above the Fermi-level. Therefore, the empty $d_{x^2-y^2}$ orbital of Pt atoms can contribute about 2e$^-$ to the $1\pi_g^*$ (formed by $p_x$ and $p_y$ orbitals) of N$_2$, making a N=N, as shown in Fig. 4(h). In turn, the N=N will favor a 2D planar structure, satisfying the Cario pentagonal tiling. Consequently, the Penta-Pt$_2$N$_4$ structure is a combined result of N=N and Cario pentagonal tiling, which is physically rooted in the coupling between $d$ orbitals of Pt and molecular orbitals of N$_2$. Following this logic, all X main group pernitrides will have a planar structure with N=N. The corresponding results for Penta-Ni$_2$N$_4$ and Penta-Pd$_2$N$_4$ are shown in the SI.

**Possible synthesis.** In the last part of our work, a possible experiment based the MBE/CVD technique is proposed to synthesize the Penta-Pt$_2$N$_4$. By using the substrate-analyzer-module in Pymatgen[65], MgF$_2$ (010) is found to be a suitable substrate for growing Penta-Pt$_2$N$_4$. Its melting point is more than 1500 K, and its lattice constant (4.69 Å) is very close to that of Penta-Pt$_2$N$_4$.

To study the energetic stability of Penta-Pt$_2$N$_4$/MgF$_2$(010), its formation energy is calculated as

$$E_{form} = (E_{Pt_2N_4/MgF_2(010)} - E_{Pt_2N_4} - E_{MgF_2(010)})/n \quad (3)$$

where $n$ is the total number of atoms, $E_{Pt_2N_4/MgF_2(010)}$, $E_{Pt_2N_4}$ and $E_{MgF_2(010)}$ are energy of Penta-$Pt_2N_4$/$MgF_2$(010), Penta-$Pt_2N_4$ and $MgF_2$(010), respectively. As shown in Fig. 4(i), the optimized distance is 2.78 Å between Penta-$Pt_2N_4$ and $MgF_2$(010), which is similar to graphene on Rh(111) substrate. The calculated $E_{form}$ is ~11 meV/atom and Bader charge analysis[66,67] shows a negligible charge transfer between Penta-$Pt_2N_4$ and $MgF_2$(010). These results indicate a weak van der Waals interaction between Penta-$Pt_2N_4$ and $MgF_2$(010), making it possible to exfoliate the synthesized Penta-$Pt_2N_4$ from substrate.

To explore the growth possibility of Penta-$Pt_2N_4$/$MgF_2$(010), we design the following reaction

$$MgF_2(010) + 2Pt + 2N_2H_4 \rightarrow Pt_2N_4/MgF_2(010) + 4H_2 \quad (4)$$

where $N_2H_4$ and bulk Pt are used as N and Pt source, respectively. The Gibbs free energy change $\Delta G$ of this reaction can be written as[68-70]

$$\Delta G = G_{Pt_2N_4/MgF_2(010)} + 4G_{H_2} - 2G_{N_2H_4} - 2G_{Pt} - G_{MgF_2(010)} \quad (5)$$

where $G$ is the Gibbs free energy of different species. Here, the DFT energy ($E$) is used to approximate Gibbs free energy of Penta-$Pt_2N_4$/$MgF_2$(010), Pt and $MgF_2$(010), since the entropy and enthalpy contribution to ΔG are negligible for solids as reported by Reuter[71]. As for gaseous molecules, the Gibbs free energy can be estimated by the following equation

$$G_{gas}(T,p) = E_{gas} + \tilde{\mu}_{gas}(T,p_0) + k_B T ln \frac{p}{p_0} \quad (6)$$

where $T$ and $p$ is temperature and pressure of the gas, respectively. $G_{gas}(T, p)$ is the Gibbs free energy, $E_{gas}$ is the DFT total energy, $p_0$ is the standard pressure ($p_0$=1 bar), and $k_B$ is the Boltzmann constant. $\tilde{\mu}_{gas}$ is the change of Gibbs free energy for gaseous molecule from 0 K to T at a constant pressure $p_0$ that can be obtained from the NIST-JANAF thermodynamics table[72,73], In order to obtain a mild synthesis condition in the experiment, such as the temperature and partial pressure ratio [$\chi$= $p(H_2)/p(N_2H_4)$], $\Delta G(T, p)$ is further derived (see SI for details). As shown in Fig. 4(j), $\Delta G$ is plotted as a function of partial pressure of $H_2$ under various partial pressure ratio at 800 K. According to the LeChatelier's principle[74], the reaction equilibrium will move to the right under lower $H_2$ pressure. The negative value region of $\Delta G$ (shadow color region) corresponds to the possible experimental growth condition. Theoretically, one can choose any point in this region. Furthermore, AIMD simulation is carried out to identify the thermal stability of Penta-$Pt_2N_4$/$MgF_2$(010) at finite temperature. As shown in Fig. S11(a)-(b), Penta-$Pt_2N_4$ is very stable on $MgF_2$(010) without notable distortion after heating 5 ps at 800 K. Therefore, $MgF_2$(010) is expected to be a good substrate for synthesizing 2D planar Penta-$Pt_2N_4$.

**Conclusion:**

In summary, combing Cario pentagonal tiling and nitrogen double bond, the first 2D Penta-$Pt_2N_4$ is theoretically predicted. The novel geometric topology and exotic chemical compositions gives excellent mechanical and electronic properties for Penta-$Pt_2N_4$, overcoming the intrinsic limitations in graphene, TMDs and black phosphorene and demonstrating an ideal 2D material for nanoelectronics. To facilitate the experiment, a possible MBE/CVD synthesis method is also proposed. We believe the great potential applications of Penta-Pt$_2$N$_4$ will stimulate more experimental works on this novel 2D material in the future.

**Methods**

Global minimal search is carried out with different unit cells of (PtN$_2$)$_n$ {n=1, 2, 3, 4} in USPEX[75] and local optimized structures are selected by comparing their average energy. During the global searching, we obtain the Penta-Pt$_2$N$_4$ and some other structures, including β-Pt$_3$N$_6$, γ-PtN$_2$ and δ-Pt$_4$N$_8$, as shown in Fig. 1(e). The β-Pt$_3$N$_6$ has a building block similar to the Penta-Pt$_2$N$_4$, but it is not fully constructed by five-membered rings. The γ-Pt$_2$N$_4$ is a highly symmetric structure constructed by both four- and six-membered rings. The δ-Pt$_4$N$_8$ has the largest unit cell and its building block is a rectangle.

DFT calculations are carried out by using the Vienna ab initio simulation package (VASP) with plane-wave basis set[76]. The projector augmented wave (PAW) method[77] is adopted in conjugation with a generalized gradient approximation (GGA)[78] of exchange-correlation function in the Perdew, Burke and Ernzerhof (PBE)[79] functional. For geometric optimization and electronic properties calculation, a plane-wave cutoff of 600 eV is used. The energy convergence criteria is $10^{-6}$ eV, and the residual force is 0.01 eV/Å. Partial occupations of eigenstates are determined by the first-order

Methfessel-Paxton smearing[80] with σ=0.05 eV. The BZ integration is carried out with 20×20×1 k-point sampling for structure optimization and 16×16×1 for electronic calculation. HSE06[81,82] is used to correct the band gap. For Penta-$Pt_2N_4$/$MgF_2$(010) calculation, van der Waals (vdW) correction proposed by Grimme (DFT-D3)[98,99] is used.

The band gap of Penta-$Pt_2N_4$ is also corrected by many-body Green's function (GW) method[83]. The GW calculations are started with the ground state wavefunction of DFT, which is implemented in Quantum Espresso package[84]. A plane-wave basis set with a kinetic energy cutoff of 110 Ry, and a norm-conserving Troullier Martins pseudopotential is used. The k-point grid sampling of 12×12×1 is used in both DFT and GW calculations. The convergence of the quasi-particle band gap is tested to be 0.1 eV, with 360 empty bands. All the GW calculations with SOC are performed by Yambo code[85].

For crystal orbital Hamilton population (COHP)[86-88] analysis, LOBSTER[89] package is used and all the absolute charge spilling is within 1%, guaranteeing the good projection.

For phonon spectrum calculation, PHONOPY[90] package is used within the density functional perturbation theory, with a plane-wave cutoff of 650 eV and the energy convergence criteria of $10^{-8}$ eV.

For mechanical property calculation, elastic constants, Young's modulus and Poisson's ratio[91-93] are calculated by PyGEC package[94] with a VASP interface.

For ab initio molecular dynamics (AIMD) simulation, a 3×3×1 supercell is used with 2×2×1 k-point sampling to reduce the lattice translational constraints. For Penta-Pt$_2$N$_4$/MgF$_2$(010), a 4×4×1 supercell is used with single Gamma point. All simulations are carried out with a Nosė-Hoover heat bath[95] at the target temperature for 5 ps with a time step of 1 fs by using canonical ensemble.

Carrier mobility calculation. In 2D system, the carrier mobility is given by the expression[96,97]

$$\mu_{2D} = \frac{e\hbar^3 C_{2D}}{k_B T m_e^* m_d (E_l^i)^2} \quad (7)$$

where $m_e^*$ is the effective mass in transport direction and $m_d$ is the average effective mass determined by $m_d = \sqrt{m_x^* m_y^*}$. $E_l^i = \Delta V_i / (\Delta l/l_0)$ is the deformation potential, where $\Delta V_i$ is energy change of the i$^{th}$ band under cell compression or dilatation, $l_0$ is the lattice constant in transport direction and $\Delta l$ is the deformation of $l_0$. $C_{2D}$ is the elastic modulus of longitudinal strain in transport direction (*x* or diagonal) for the longitudinal acoustic wave. It can be derived from $(E-E_0)/S_0 = C_{2D} (\Delta l/l_0)^2/2$, where $E$ is the total energy and $S_0$ is the lattice volume at equilibrium for a 2D system that can be calculated by PyGEC. In carrier mobility calculations, the deformation potential ($E_1$) for hole and electron is derived from a linear fitting to the energy of CBM and

VBM versus the lattice compression or dilatation with a step of 0.33%, respectively

*Rev. Lett.* **96,** 155501 (2006).

33. Niwa, K. *et al.* Discovery of the last remaining binary platinum-group pernitride RuN$_2$. *Chem. Eur. J.* **20,** 13885 (2014).

34. Bhadram, V. S., Kim, D. Y. & Strobel, T. A. High-pressure synthesis and characterization of incompressible titanium pernitride. *Chem. Mater.* **28,** 1616-1620 (2016).

35. Zhang, M. *et al.* Electronic bonding analyses and mechanical strengths of incompressible tetragonal transition metal dinitrides TMN$_2$ (TM=Ti, Zr, and Hf). *Sci. Rep.* **6,** 36911 (2016).

36. Schneider, S. B., Frankovsky, R. & Schnick, W. High-pressure synthesis and characterization of the alkali diazenide. Li$_2$N$_2$ *Angew. Chem.* **124,** 1909-1911 (2012).

37. Schneider. S. B. *et al.* High-pressure synthesis and characterization of Li$_2$Ca$_3$[N$_2$]$_3$-an uncommon metallic diazenide with [N$_2$]$^{2-}$ ions. *J. Am. Chem. Soc.* **135,** 16668-16679 (2013).

38. Niwa, K. et al High pressure synthesis of marcasite-type rhodium pernitride. *Inorg. Chem.* **53,** 697-699 (2014).

39. Wang, Z. *et al.* Prediction and characterization of the marcasite phase of iron pernitride under high pressure. *J. Alloy. Compd.* **702,** 132-137 (2017).

40. Vajenine, G. V. *et al.* Preparation, crystal structure and properties of barium pernitride, BaN$_2$. *Inorg. Chem.* **40,** 4866-4870 (2001).

41. Schneider, S. B., Frankovsky, R. & Schnick, W. Synthesis of alkaline earth diazenides M$_{AE}$N$_2$ (M$_{AE}$=Ca, Sr, Ba) by control thermal decomposition of azides
20

| Molecules | $CO_2$ | $H_2$ | $N_2$ | $O_2$ | $H_2O$ |
|-----------|--------|-------|-------|-------|--------|
| $E_a$ (meV) | −176 | −70 | −123 | −324 | −146 |

Table 1 | Adsorption energy of different molecules on Penta-$Pt_2N_4$.

| Carrier Type | $m^*$ ($m_0$) | | $E_l$ (eV) | | $\mu_{2D}$ ($10^4 cm^2 \cdot V^{-1} \cdot s^{-1}$) | |
|---|---|---|---|---|---|---|
| | $x$ | Diagonal | $x$ | Diagonal | $x$ | Diagonal |
| e | 0.362 | 0.485 | −0.566 | −1.113 | 11.418 | 1.502 |
| h | 0.071 | 0.472 | −0.860 | −1.714 | 9.360 | 0.529 |

Table 2 | Effective masses ($m^*$), deformation potential ($E_l$) and carrier mobility ($\mu_{2D}$) of Penta-$Pt_2N_4$ along $x$ and diagonal direction. The results are calculated at T=298 K on PBE+SOC level.

| Properties | Penta-$Pt_2N_4$ | graphene | $MoS_2$ | black phosphorene |
|---|---|---|---|---|
| Band gap (eV) | / | 0 (e) | 1.8 (e) | 0.6 (e) |
| | 1.51 | 0 (t) | 1.78 (t) | 1.51 (t) |
| Carrier mobility ($10^4 cm^2 \cdot V^{-1} \cdot s^{-1}$) | / | 20 (e) | 0.02 (e) | 0.1 (e) |
| | 11 | 30 (t) | 0.03 (t) | 1.0 (t) |
| Air stable | Yes | Yes | Yes | No |
| Young's modulus (TPa) | / | 1.0 (e) | 0.33 (e) | / |
| | 0.70 | 1.0 (t) | 0.25 (t) | 0.17 (t) |
| Poisson's ratio | 0.47 | 0.19 (t) | 0.21 (t) | 0.93 (t) |
| References | This work | 7-8, 51-55 | 9-10, 61-64 | 11-15, 56-60 |

Table 3 | Summary of electronic and mechanical properties of Penta-$Pt_2N_4$ compared to those of graphene, $MoS_2$ and black phosphorene[*]. [*]Experimental properties of monolayer black phosphorene haven't been reported yet, so only few layer results are listed. The label "e" and "t" denote experimental and theoretical value, respectively.

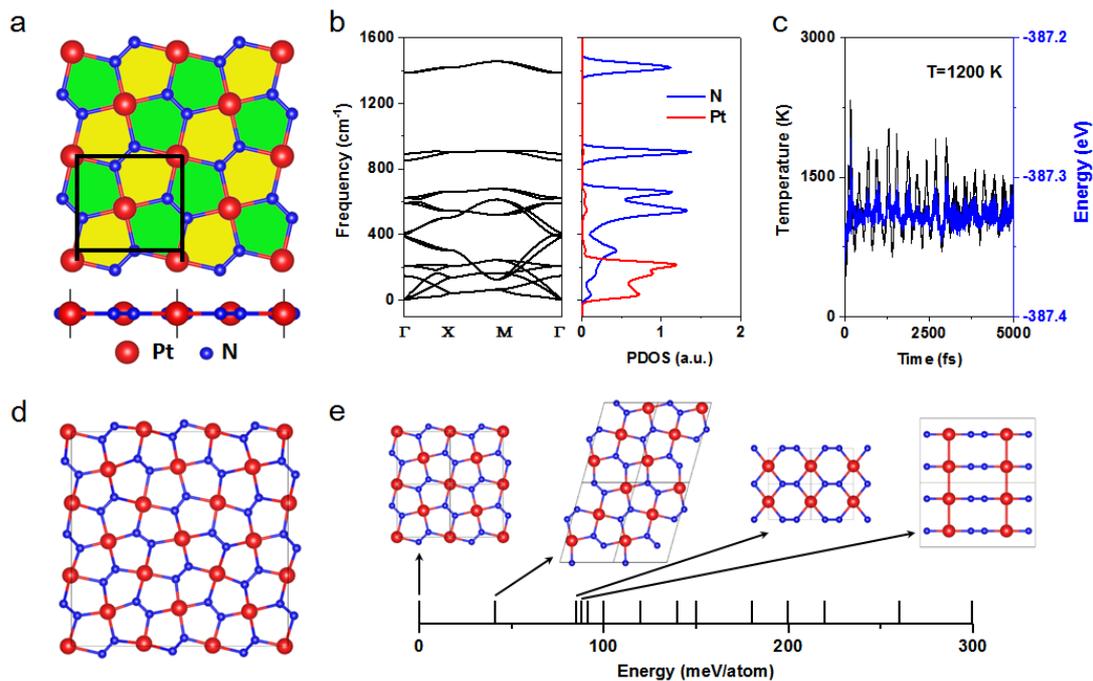

**Figure 1 | Atomic structure and stability of Penta-Pt$_2$N$_4$.** (a) Cario pentagonal tiling and crystal structure. (b) Phonon spectrum and partial density of state. (c) Fluctuation of temperature and total energy at 1200 K in AIMD simulation. (d) Top view of AIMD structure taken from the last snapshot in (c). (e) Configurational energy spectrum of (PtN$_2$)$_n$ (n=1, 2, 3, 4). Penta-Pt$_2$N$_4$ is set as the reference energy.

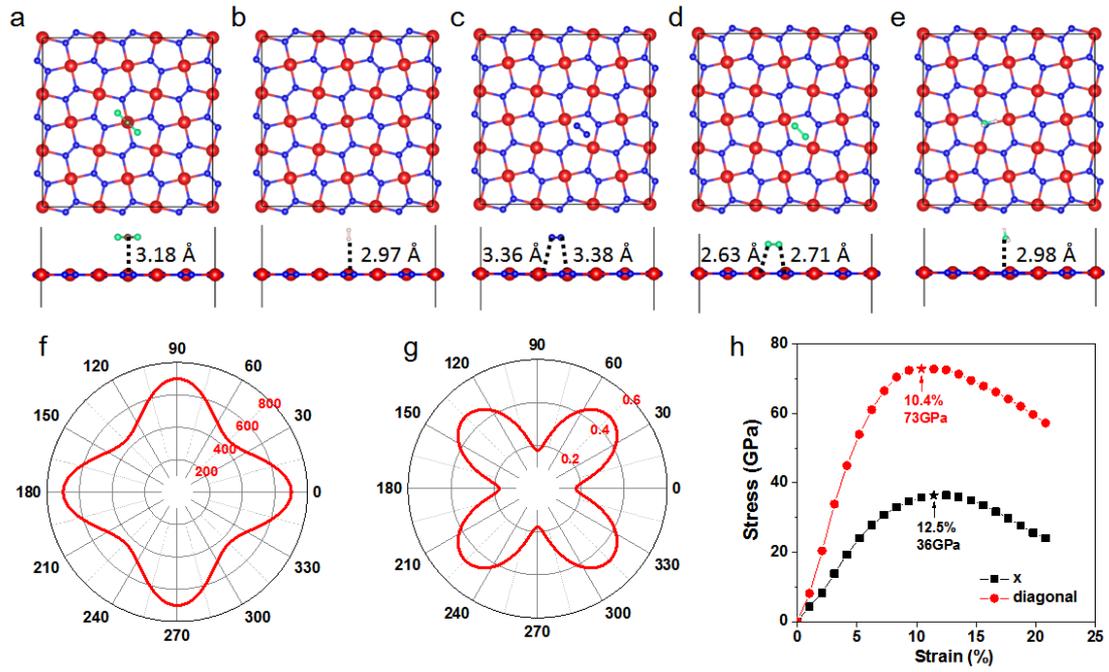

**Figure 2 | Chemical stability and mechanical properties of Penta-Pt$_2$N$_4$.** (a)-(e) Molecular adsorption structure and adsorption distance for $CO_2$, $H_2$, $N_2$, $O_2$ and $H_2O$, respectively. (f) Angular-dependent Young's modulus. (g) Angular-dependent Poisson's ratio. (h) The strain-stress relation in *x* and diagonal directions.

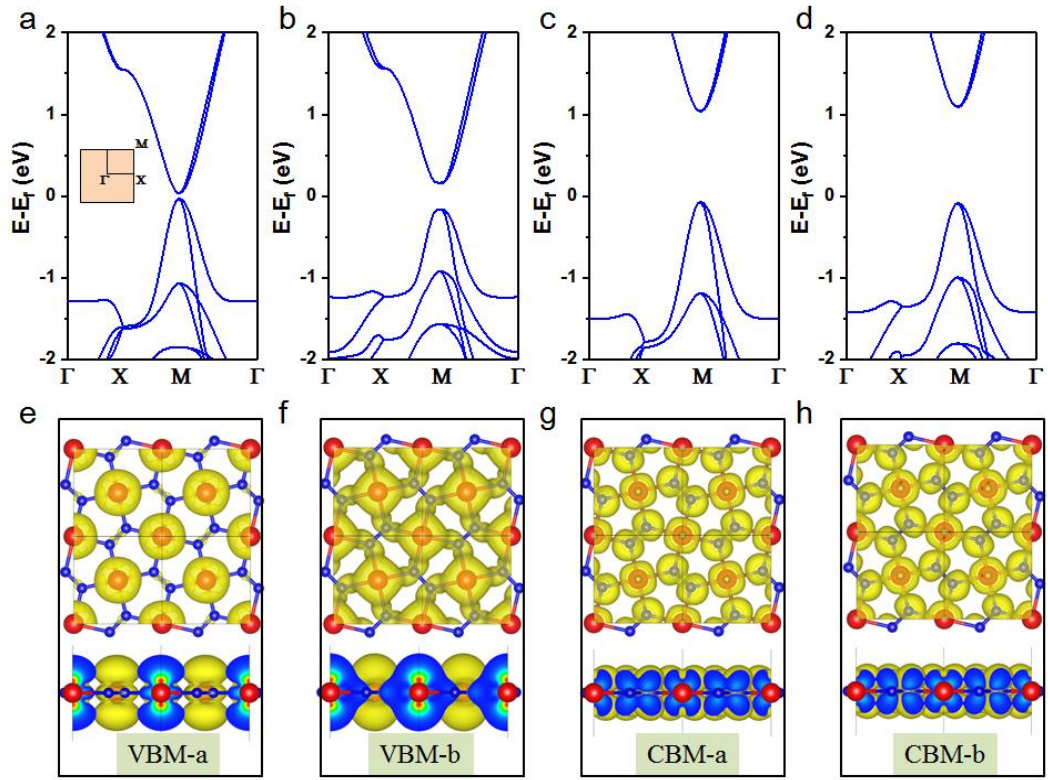

**Figure 3 | Electronic properties of Penta-Pt$_2$N$_4$.** (a) PBE band structure without SOC. Inset is the first BZ and high symmetric k points. (b) PBE band structure with SOC. (c) HSE band structure without SOC. (d) HSE band structure with SOC. (e)-(h) Charge densities at valence-band maximum (VBM) and conduction-band minimum (CBM) in (a). VBM and CBM are 2-fold degenerate, so both charge densities are displayed.

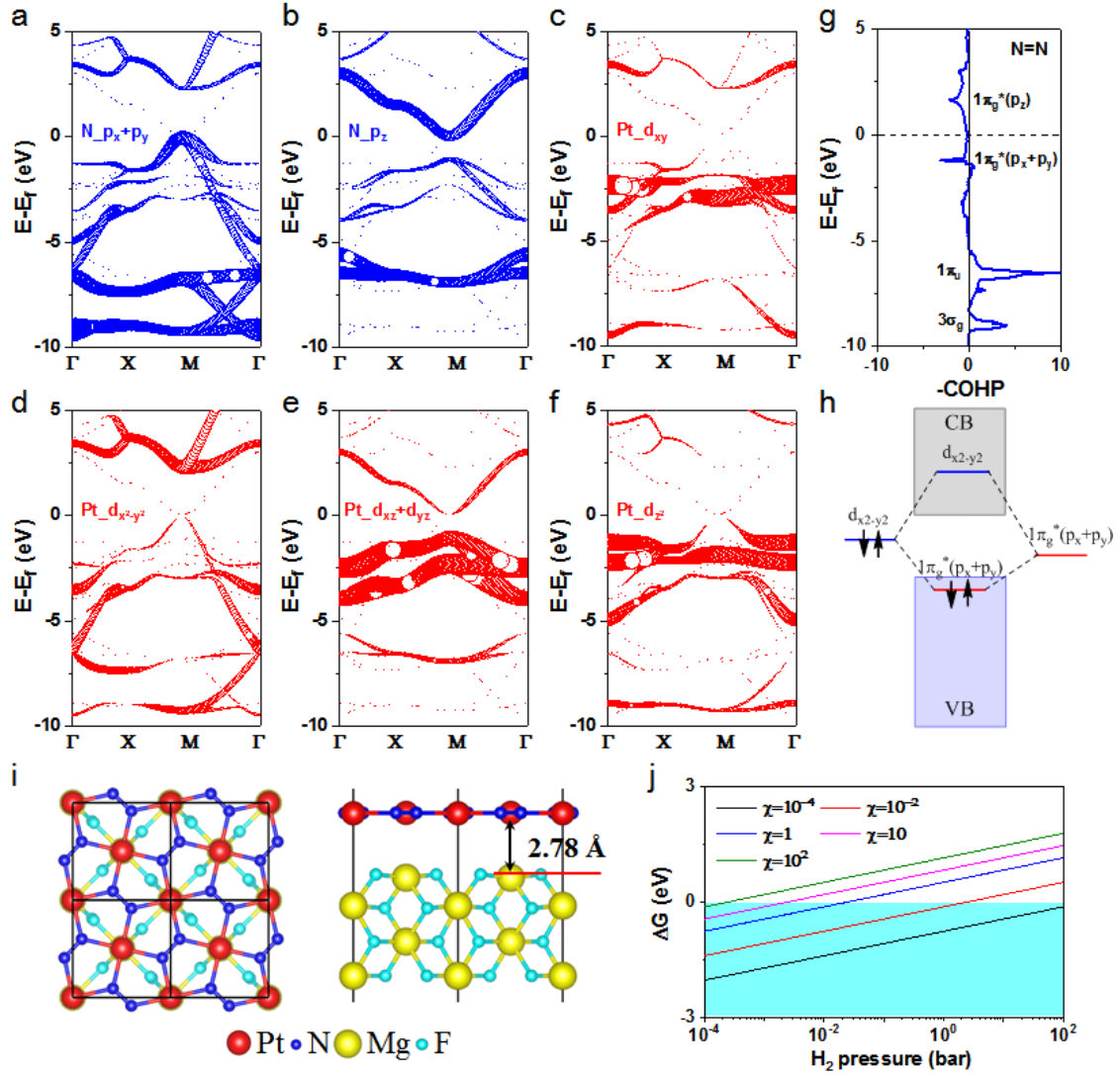

**Figure 4 | Nature of N=N and possible synthesis of Penta-Pt$_2$N$_4$.** (a)-(f) Orbital projected PBE band structures. (g) COHP analysis of N=N. (h) Schematic diagram of orbital coupling between Pt and N$_2$. (i) Top and side view of Penta-Pt$_2$N$_4$/MgF$_2$(010). (j) Gibbs free energy change ($\Delta G$) versus gas-phase H$_2$ pressure at different partial pressure ratio $\chi$.